\documentclass[11pt]{article}

\usepackage{graphicx}
\usepackage{amsmath}
\usepackage{dcolumn}
\usepackage{bm}
\usepackage{slashed}
\usepackage[bookmarks,bookmarksnumbered,colorlinks=true,anchorcolor=blue,
linkcolor=blue,urlcolor=blue,citecolor=blue,breaklinks=true]{hyperref}
\usepackage[utf8]{inputenc}
\usepackage{authblk}
\usepackage[numbers,sort&compress]{natbib}
\usepackage{color}
\usepackage{ulem}

\newcommand{\be}{\begin{equation}}
\newcommand{\ee}{\end{equation}}
\newcommand{\ba}{\begin{eqnarray}}
\newcommand{\ea}{\end{eqnarray}}
\def\bea{\begin{eqnarray}}
\def\eea{\end{eqnarray}}

\newcommand{\gsim}{\mathrel{\hbox{\rlap{\lower.55ex \hbox {$\sim$}}
                   \kern-.3em \raise.4ex \hbox{$>$}}}}
\newcommand{\lsim}{\mathrel{\hbox{\rlap{\lower.55ex \hbox {$\sim$}}
                   \kern-.3em \raise.4ex \hbox{$<$}}}}

\def\del{\partial}

\def\roughly#1{\mathrel{\raise.3ex\hbox{$#1$\kern-.75em%
\lower1ex\hbox{$\sim$}}}}
\def\lsim{\roughly<}
\def\gsim{\roughly>}

\def\({\left(}
\def\){\right)}
\def\[{\left[}
\def\]{\right]}
\def\<{\langle}
\def\>{\rangle}

\def\L{{\Lambda}}

\def\m{{\mu}}
\def\n{{\nu}}
\def\r{{\rho}}
\def\s{{\sigma}}

\def\t{{\tau}}

\usepackage{xcolor}

\newcommand{\red}[1]{{\color{red} #1 \color{black}}}

\title{\bf {Notes on the double Wick rotated BTZ black hole}}

\author[1]{Hua-Yu Dai\thanks{daihuayu@cuhk.edu.cn}}
\author[2]{Xi-Hao Fang\thanks{fangxh9@mail2.sysu.edu.cn}}
\author[3]{Mitsutoshi Fujita\thanks{fujitamitsutoshi@usc.edu.cn}}
\affil[1]{School of Science and Engineering, The Chinese University of Hong Kong (Shenzhen), Shenzhen, 518172, China}
\affil[2]{Beijing Institute of Mathematical Sciences and Applications (BIMSA), Huaibei Town, Huairou Dis-trict, Beijing 101408, China}
\affil[3]{School of Nuclear Science and Technology, University of South China, Hengyang 421001, China}

\date{}

\begin{document}

\maketitle

\begin{abstract}
{We analyze the double Wick rotated BTZ black hole with the Euclidean signature, which is a Riemannian manifold. We calculate thermodynamics, total energy of spacetime, and holographic two point functions in the double Wick rotated background. Results agree with those of a rotating BTZ with the same periodicity.  }

\end{abstract}
{}

\newpage

\allowdisplaybreaks

\flushbottom

\section{Introduction}
The BTZ (Banados-Teitelboim-Zanellit) black hole~\cite{Banados:1992wn,Banados:1992gq} is a (2+1) dimensional rotating black hole, which is specified by its mass, angular momentum, and (optionally) the electric charge like the Kerr black hole. A rotating BTZ black hole is not static but stationary. Unlike a Schwarzschild black hole, it does not exhibit curvature singularities at the origin in the absence of matter couplings, while the surface in the origin becomes singular in the causal structure.~\footnote{ It can be derived from the Poincare AdS coordinates through a specific coordinate transformation.} {The universal exponent of the black hole formation, which is known as the Choptuik scaling parameter, was obtained for the BTZ black hole and was related with the imaginary part of quasi-normal modes \cite{Birmingham:2001hc}.} The quantum correction for the path integral has been studied in \cite{Carlip:1994gc,Carlip:1992us}. Additionally, the quantum corrections of temperature and entropy were obtained in \cite{Ghosh:1994nz}. Quantum entanglement in a BTZ black hole was analyzed in ~\cite{Hubeny:2007xt}-\cite{Boldis:2021snw}. {Entanglement entropy in CFTs dual to extremal BTZ has the right movers for the thermal form with Frolov-Thorne temperature~\cite{Caputa:2013lfa}.}
   
{The double Wick rotated BTZ black hole was proposed  in~\cite{Fujita:2022zvz} to analyze dual CFT, in which the periodicity of Euclidean time and a space direction is interchanged. This metric has problems about causality in the Lorentzian frame. Analyzing a Killing vector to see the causal structure, a closed timelike curve  exists inside the double Wick rotated metric. It appears behind the Cauchy horizon (or a lightlike boundary)~\cite{Banados:1992gq}. It is crucial to examine how the law of physics allows a closed time-like curve.~\cite{Friedman:1990xc}} The double Wick rotated BTZ black hole also has negative energy, dual to Casimir energy in the CFT side. Covariant entanglement entropy in this holographic model factorizes into the left and right sectors. The factorization arises regardless of the double Wick rotation of spacetimes because the (double Wick rotated) BTZ black hole has a torus-like structure~\cite{Hubeny:2007xt}. In contrast, the phenomenon of factorization disappears in the cylinder spacetime, where one of the periodicities go to infinity.
 
The analytically continued Euclidean metric of the rotating BTZ black hole is written as \eqref{MET1}, where its angular momentum is also analytically continued. At that case, a similarity emerges between a rotating BTZ and the double Wick rotated background. Due to the analytic continuation ($u_-\to -i\tilde{u}_-$), the double Wick rotated background has the black hole horizon and Hawking temperature. It was pointed out in~\cite{Fujita:2022zvz} that the holographic stress tensor of both theories becomes the same form if the periodicity of both a rotating BTZ and the double Wick rotated background is identified in the Euclidean frame. Because both theories are dual to a CFT with the same peridocity, the motivation of this paper is to apply this equivalence for other observables. 

In this paper, we analyze thermodynamics and correlation functions of the double Wick rotated background. We derive total energy of spacetime, entropy, and free energy from the double Wick rotated background. We confirm an agreement of thermodynamics between two backgrounds. Note that one must perform an analytic continuation  ($u_-\to -i\tilde{u}_-$) to compare two black hole spacetimes. 
{The central charge of dual CFT is examined. The central charge only depends solely on the gravitational coupling and the AdS radius in double Wick rotation case. Therefore, the central charge does not depend on mass and angular momentum of the black hole. One can figure it out that twice analytic continuation conicidently maintain the central charge itself.} The propagator is also analyzed by using the geodesic approximation. It requires infinite sums because of periodicities. 

This paper is organized as follows. In section \ref{EBTZ}, we revisit the Euclidean rotating BTZ black hole. We derive thermodynamics, total energy of the spacetime, central charges, and holographic 2-point functions.  In section \ref{DWR1}, we analyze the double Wick rotated BTZ background and show similarity to the Euclidean rotating BTZ black hole, discuss both  thermodynamics and holographic charasteristics.  Last in section \ref{DSC}, we summarize our discussion of double wick rotation black hole and consider the potential future research directions. Appendix \ref{ADM}  review the ADM decomposition of spacetime of BTZ and DW (double Wick) BTZ respectively.

\section{The Euclidean rotating BTZ black hole}\label{EBTZ}
In this section, we revisit the Euclidean rotating BTZ black hole \cite{Banados:1992wn,Banados:1992gq,Carlip:1995qv}. We derive thermodynamic variables and total energy of the spacetime, focusing on the case with time translation invariance. We begin with the following action: 
\ba\label{IEU}
I_{E}[g]=-\frac{1}{16\pi G_{3}}\int d^{3}x\sqrt{g}\(R-2\L\)-\frac{1}{8\pi G_{3}}\int d^{2}x\sqrt{h}(K-\kappa),
\ea
where $\Lambda =-1/L^2$ is the cosmological constant, $h_{ij}$ is the induced metric on the AdS boundary, $K$ is the extrinsic curvature. $\kappa= 1/L$ is the coefficient of a counter term to render the action finite~\cite{Huang:2021eby}.

{The Euclidean rotating BTZ black hole is a Riemannian manifold and a solution of the Einstein equation derived from \eqref{IEU} as follows:
\ba\label{MET1}
&ds^2=L^2\Big[ u^2 \Big(dx -\dfrac{u_+\tilde{u}_-}{u^2}d\tau_E \Big)^{{2}} +\dfrac{u^2 du^2}{(u^2-u_+^2)(u^2+\tilde{u}_-^2)}+\dfrac{(u^2+\tilde{u}^2_-)(u^2-u_+^2)}{{u}^2}d\tau_E^2\Big], \nonumber \\
&8G_3 M=u_+^2-\tilde{u}_-^2,\quad J_E=\dfrac{Lu_+\tilde{u}_-}{4G_3}.
\ea
where $x \sim x+2\pi$ and $L$ is the AdS radius. Note that $M$ and $G_3^{-1}$ have the dimension 1 and $J_E$ has the dimension 0. This metric is obtained via the analytic continuation of the rotating BTZ black hole: $t\to -i \tau_E$ and $J\to -i J_E$ ($u_-\to -i\tilde{u}_-$). The continuation of $J$ to imaginary values is necessary to have a Riemannian metic. Actually, this metric is positive definite: $g_{\mu\nu}=\delta_{ab}e^a_{\mu}e^b_{\nu}$ and $g_{\mu\nu}e^{\mu}_a e^{\nu}_b=\delta_{ab}$, where $e^{\mu}_a$ is a tangent frame field and $\delta_{ab}$ is a Kronecker symbol.} {According to the transformation of $u_-$, the lapse is changed into 
\ba\label{LAP35}
N=\dfrac{L}{u}\sqrt{(u^2+\tilde{u}_-^2)(u^2-u_+^2)}.
\ea
See appendix~\ref{ADM} for details on the shift. The metric \eqref{MET1} exhibits periodicity $(\tau_E,x)\sim (\tau_E+\eta , x+\zeta)$~\cite{Carlip:1994gc}:
\ba\label{PER2}
\eta =\dfrac{2\pi {u}_+}{u_+^2+\tilde{u}_-^2},\quad \zeta =\dfrac{2\pi \tilde{u}_-}{u_+^2+\tilde{u}_-^2}.
\ea
Temperature and chemical potential are defined as~\cite{Banados:1992wn,Banados:1992gq}
\ba
\beta=\eta L =\dfrac{2\pi u_+L}{u_+^2+\tilde{u}_-^2},\quad \Omega=-\dfrac{4 G_3 J}{u_+^2L^2}=-\dfrac{u_-}{u_+ L}=\dfrac{i \tilde{u}_-}{u_+ L}.
\ea
 The definition of chemical potential is with respect to the shift: $\Omega =-N^x (u_+)/L$, where the shift is defined as $N^x(u)=u^+u^-/u^2$ (See appendix \ref{ADM}). Note that the sign of the shift changes due to parity symmetry $x\to -x$. Note that temperature (or periodicity) and the chemical potential are finite for non-zero values of $u_+$ and $\tilde{u}_-$. Due to the analytic continuation, the zero temperature (extremal) limit $u_+=u_-$ can not be taken. 

We proceed to investigate the thermodynamics of the black hole. Energy of the Euclidean rotating BTZ black hole is obtained by computing total energy of the spacetime~\cite{Brown:1992br}. Total energy is defined in a background that possesses time translation symmetry. 
\ba\label{TOT39}
E=\dfrac{1}{8\pi G_3}\int dx N \sqrt{\sigma} (K-K_0).
\ea
$K_0$ is evaluated on reference spacetime with $u_+=\tilde{u}_-=0$. Considering a spatial slice of constant $t$, $u$ is also fixed. One then has the area $A=L uV_x$, where  $V_x$ is the volume of $x$ direction, namely, $V_x=2\pi$. The lapse function is defined in \eqref{LAP35}. 
The trace of the extrinsic curvature is evaluated at fixed $u$ and $t$. Consequently, \eqref{TOT39} is computed as 
\ba
E=\dfrac{(u_+^2-\tilde{u}_-^2)V_xL}{16\pi G_3}=\dfrac{(u_+^2-\tilde{u}_-^2)L}{8 G_3}.
\ea
After restoring dimensions ($E= \tilde{E}L$ and $P=\tilde{P}L$), energy becomes
\ba\label{ENE40}
\tilde{E}=M=\dfrac{u_+^2-\tilde{u}_-^2}{8\pi G_3}.
\ea
It matches the mass of the black hole as anticipated. 

Energy and momentum can also be computed from the holographic stress tensor with counter terms \cite{deHaro:2000vlm, Balasubramanian:1999re}. See~\cite{Fujita:2022zvz,He:2014lfa}. One needs to perform analytic continuation $\tau_E=it$ to compute $\langle T_{00} \rangle$ and $\langle T_{0x}\rangle$. The resulting expressions for  
 energy and momentum become 
\ba
\tilde{E}=\int dx \langle T_{00} \rangle=M,\quad \tilde{P}=\dfrac{1}{L}\int dx \langle T_{0x} \rangle=\dfrac{J}{L}.
\ea

Considering the constant time slice and the constant radial slice, the Bekenstein-Hawking entropy is proportional to the horizon area as follows:
\ba\label{SEN1}
s=\dfrac{V_x u_+ L}{4G_3}.
\ea
This corresponds to $1/4$ of the horizon size in Planck units.

Moreover, we evaluate the grand canonical partition function for the Euclidean path integral in the steepest descent (classical) approximation $Z\sim e^{-I_E}[g]$. In accordance with statistical mechanics and the AdS/CFT correspondence, free energy is temperature times the value of the Euclidean action on the Euclidean continuation of the black hole. Then what we need to calculate is the Euclidean action \eqref{IEU}. As for the first term of \eqref{IEU}, we have
\be\label{IEU1t}
\begin{aligned}
-\frac{1}{16\pi G_{3}}\int d^{3}x\sqrt{g}\(R-2\L\)&=-\frac{1}{16\pi G_{3}}\cdot 4\L\int d^{3}x\sqrt{g}\\
&=\frac{1}{4\pi G_{3}L^{2}}\int dxd\tau_{E}\int du L^{3}u\\
&=\frac{\eta L}{4G_{3}}(u^{2}_{\infty}-u_+^2).
\end{aligned}
\ee
Here we have utilized the relations, $R=6\L$, $\L=-1/L^{2}$ and $\sqrt{g}=L^{3}u$, in metric \eqref{MET1} to derive the second line. Furthermore, we evaluate the integral $\int dxd\tau_{E}=2\pi\eta$ to derive the final expression. 

As for computing the extrinsic curvature scalar with radial slicing, we define the unit normal vector as 
\be
n_{\mu}=\frac{\nabla_{\mu}u}{\sqrt{g^{\mu\nu}\nabla_{\mu}u\nabla_{\nu}u}}.
\ee
Then we calculate the extrinsic curvature scalar as follows:
\be
\begin{aligned}
K\equiv\nabla_{\mu}n^{\mu}&=\frac{1}{\sqrt{g}}\del_{\mu}\(\sqrt{g}g^{\mu\nu}n_{\nu}\)\\
&=\frac{1}{\sqrt{g}}\del_{u}\(\sqrt{g}\sqrt{g^{uu}}\)\\
&=\frac{1}{2Lu\sqrt{A(u)}}\frac{dA(u)}{du},
\end{aligned}
\ee
where we introduce the notation $A(u)\equiv\(u^{2}+\tilde{u}_{-}^{2}\)\(u^{2}-u_{+}^{2}\)$ from the metric \eqref{MET1}. Therefore, the second term in \eqref{IEU} can be derived as
\be
\begin{aligned}
-\frac{1}{8\pi G_{3}}\int_{u=u_{\infty}} d^{2}x\sqrt{h}K&=-\frac{1}{8\pi G_{3}}\int_{u=u_{\infty}} dxd\tau_{E}L^{2}\sqrt{A(u)}\cdot\frac{1}{2Lu\sqrt{A(u)}}\frac{dA(u)}{du}\\
&=-\frac{\eta L}{4G_{3}}\left[2u^{2}_{\infty}+\(\tilde{u}_{-}^{2}-u_{+}^{2}\)\right].
\end{aligned}
\ee
Furthermore, the third term in \eqref{IEU} is
\be
\begin{aligned}
\frac{1}{8\pi G_{3}}\int_{u=u_{\infty}} d^{2}x\sqrt{h}\frac{1}{L}&=\frac{1}{8\pi G_{3}}\int_{u=u_{\infty}} dxd\tau_{E}L^{2}\sqrt{A(u)}\cdot\frac{1}{L}\\
&=\frac{\eta L}{4G_{3}}\sqrt{u^{4}+\(\tilde{u}_{-}^{2}-u_{+}^{2}\)u^{2}-\tilde{u}_{-}^{2}u_{+}^{2}}\\
&=\frac{\eta L}{4G_{3}}u^{2}\(1+\frac{\tilde{u}_{-}^{2}-u_{+}^{2}}{2u^{2}}+\mathcal{O}(u^{-2})\)\\
&\approx\frac{\eta L}{4G_{3}}\left[u^{2}_{\infty}+\frac{1}{2}\(\tilde{u}_{-}^{2}-u_{+}^{2}\)\right].
\end{aligned}
\ee
Combining these, we obtain free energy as follows:
\ba
F=\dfrac{I_E(g)}{\eta L}=-\dfrac{1}{8G_3}(u_+^2+\tilde{u}_-^2),
\ea
where we have substituted the temperature via $T=\frac{1}{\eta L}$. 
Free energy satisfies the following thermodynamic relation:
\ba\label{FEE43}
&\tilde{E}-Ts -\Omega J =\dfrac{u_+^2-\tilde{u}_-^2}{8G_3}-\dfrac{u_+^2+\tilde{u}_-^2}{2\pi u_+L}\dfrac{2\pi u_+L}{4G_3}+\dfrac{i\tilde{u}_-}{u_+L}\dfrac{-i L u_+ \tilde{u}_-}{4G_3} \nonumber \\
&=-\dfrac{u_+^2+\tilde{u}_-^2}{8G_3}=F.
\ea
Comparing this with the Euler relation $\tilde{E}=Ts-PV_x +\mu Q$, free energy is written as the product of  pressure and volume as follows: $F=-PV_x$. 

{A nonzero trace of the holographic stress tensor for an asymptotic AdS metric gives the Weyl anomaly. The central charge can be obtained from the coefficient of the Weyl anomaly as follows~\cite{deHaro:2000vlm, Balasubramanian:1999re}:
\ba
\langle T^{\mu}{}_{\mu}\rangle =-\dfrac{c}{24\pi}R, 
\ea
where $c=3L/(2G_3)$.  It's originally obtained by Brown and Henneaux~\cite{Brown:1986nw}.

Note that the central charge does not depend on the mass and angular momentum of the rotating BTZ black hole. Consequently, an analytic continuation does not alter the central charge. 
}

\subsection{The propagator in the Euclidean rotating BTZ}
In this section, we employ the geodesic approximation to derive holographic two point function in the Euclidean rotating BTZ metric. This section serves as a review of the work presented in~\cite{Louko:2000tp}.

{The Euclidean rotating BTZ is a Riemannian metric with constant negative curvature and is locally isometric to hyperbolic 3 space $H^3$. We consider the Euclidean AdS. The following hyperboloid represents $H^3$:
\ba
-(T_1)^2+(T_E)^2+(X_1)^2+(X_2)^2=-1,
\ea
where an analytic continuation is performed $T_E=-iT_2$ because it is the Euclidean signature. 
Note the minus sign in the right-hand side. Restricting to this submanifold, the metric is written as 
\ba\label{MET1t}
ds^2=-(dT^1)^2+(dT_E)^2+(dX^1)^2+(dX^2)^2.
\ea
The Euclidean rotating BTZ is obtained via the following coordinate transformation:}
\ba\label{GEN29}
&T_1=\sqrt{\beta_1}\cosh (\tilde{u}_-\tau_E{+} u_+ x), \nonumber \\
&T_E=-\sqrt{\beta_1 -1}\sin (u_+ \tau_E {-}\tilde{u}_- x), \nonumber \\
&X_1=-\sqrt{\beta_1}\sinh ({\tilde{u}_-} \tau_E {+}u_+ x), \nonumber \\
&X_2=\sqrt{\beta_1 -1}\cos (u_+ \tau_E\red{-}{\tilde{u}_-} x).
\ea 
where $\beta_1 =(u^2+\tilde{u}_-^2)/(\tilde{u}^2+u_+^2)$. $u>u_+$. The periodicity of $x$ is $x\sim x+2\pi$ due to the rotating BTZ quotient construction. For $\tilde{u}_- =0$, the transformation gives a BTZ black hole without angular momentum. Substituting \eqref{GEN29} into \eqref{MET1t}, one obtains the metric of the Euclidean BTZ \eqref{MET1}.~\footnote{This follows with a sign flip of $\tilde{u}_-$ because of the convention.} 

We consider two points $(\tau_E^1, x^1,u)$ and $(\tau_E^2,x^2,u)$ in the bulk. And consider the chordal distance of geodesic connecting two points. Substituting \eqref{GEN29}, the chordal distance of the geodesic is transformed by 
{\ba
&-(T_1-T_1')^2+(T_E-T_E')^2+(X_1-X_1')^2+(X_2-X_2')^2 \nonumber \\
&=\dfrac{-2}{u_+^2+\tilde{u}_-^2}\Big(\tilde{u}_-^2+u_+^2 +(u^2-u_+^2)\cos (u_+ \Delta \tau_E-\tilde{u}_-\Delta x)  \nonumber \\
&-(u^2+\tilde{u}_-^2) \cosh (\tilde{u}_- \Delta \tau_E +u_+\Delta x) \Big),
\ea}
where $\Delta \tau_E =\tau_E^1 -\tau_E^2$ and $\Delta x = x^1-x^2$. In the large $u$ limit, the chordal distance of geodesic becomes
\ba\label{CHO1}
-\dfrac{2u^2}{\tilde{u}_-^2+u_+^2}(\cos (u_+ \Delta \tau_E-\tilde{u}_-\Delta x)-\cosh (\tilde{u}_-\Delta \tau_E+u_+ \Delta x  ))
\ea 
Because $\cosh (x)\ge 1$, this geodesic remains spacelike for sufficiently large $u$. 

The chordal distance is connected to the proper length of geodesic as follows:
\ba\label{PRO1}
\sinh^2 (L/2)=\dfrac{D}{4}.
\ea
This relationship allows us to employ the geodesic approximation of the holographic correlation function. {Recall that the BTZ quotient construction means the identification $x\sim x+2\pi$, meaning $x_1$ belongs to the equivalent class of $x_2$. Consequently, the holographic correlation function becomes
\ba\label{HLO1}
&\langle \phi(x_1)\phi(x_2)\rangle =\int dP e^{i \Delta L(P)}=\sum_{n}e^{-\Delta L_n(x_1,x_2)} \nonumber \\ 
&=\sum_n \dfrac{(2u^2)^{-\Delta}(u_+^2+\tilde{u}_-^2)^{\Delta}}{(\cos (u_+ \Delta \tau_E-\tilde{u}_-(\Delta x+2\pi n))-\cosh (\tilde{u}_-\Delta \tau_E+u_+ (\Delta x +2\pi n) )))^{\Delta}}
\ea
After eliminating the factor of $u$, the dominant contribution in the Green's function is obtained as follows:}
\ba
\sum_n \dfrac{({u_+^2+\tilde{u}_-^2})^{\Delta}}{(\cos (u_+ \Delta \tau_E-\tilde{u}_-(\Delta x+2\pi n))-\cosh (\tilde{u}_-\Delta \tau_E+u_+ (\Delta x +2\pi n) )))^{\Delta}}
\ea

\section{{The double Wick rotated BTZ black hole}}\label{DWR1}
{We consider the double Wick rotated BTZ black hole with angular momentum~\cite{Fujita:2022zvz}. We do the computation in the Euclidean signature. This is achieved by the double Wick rotation $\tau_E \leftrightarrow x$. Exchanging $u_+$ with $\tilde{u}_-$, furthermore, we obtain the following metric:
\ba\label{MET2}
ds^2_{DW}=L^2 \Big[u^2\Big(\dfrac{\tilde{u}_-u_+}{u^2}dx-d\tau_E\Big)^2+\dfrac{(u^2-\tilde{u}_-^2)(u^2+u_+^2)}{u^2}dx^2+\dfrac{u^2du^2}{(u^2-\tilde{u}_-^2)(u^2+u_+^2)}  \Big].
\ea
where $x\sim x+2\pi$. This is a solution of the Einstein equation derived from \eqref{IEU}. The double Wick rotated metric is also positive definite and then it is  Riemannian: $g_{\mu\nu}=\delta_{ab}e^a_{\mu}e^b_{\nu}$ where $\delta_{ab}$ is Kronecker symbol.}  
{As a result of the exchange of $u_+$, the lapse becomes
\ba\label{LAP68}
N=L\sqrt{\dfrac{(u^2-\tilde{u}_-^2)(u^2+u_+^2)}{u^2-\tilde{u}^2_-+u_+^2}}.
\ea}
The black hole horizon is located at $u=\tilde{u}_-$. As is analogous to the rotating BTZ, the double Wick rotated metric exhibits constant negative curvature and locally becomes equivalent to hyperbolic three space. 
The double Wick rotated metric is parametrized by the coordinate transformation from the Poincare AdS as follows: 
\ba
&\bar{X}=\sqrt{\delta}\cos (u_+\tau_E +\tilde{u}_-x)\exp (-u_+x +\tilde{u}_- \tau_E), \nonumber \\
&\bar{Y}=\sqrt{\delta}\sin (u_+\tau_E +\tilde{u}_-x)\exp (-u_+x +\tilde{u}_- \tau_E), \\
&z=\sqrt{1-\delta}\exp (-u_+x +\tilde{u}_- \tau_E),
\ea
where $\delta =(u^2-\tilde{u}_-^2)/(u^2+u_+^2)$. A spherical coordinate is given by $\phi_1=u_+\tau_E +\tilde{u}_-x$. The AdS boundary $u\to\infty$ corresponds to $z= 0$ and the near horizon limit $u\to \tilde{u}_-$ corresponds to the 
$z$ axis at $(\bar{X},\bar{Y})\to (0,0)$. The double Wick rotated metric is rewritten as the following Poincare AdS:
\ba
ds^2_{DW}=\dfrac{L^2}{z^2}(d\bar{X}^2+d\bar{Y}^2+dz^2).
\ea

The periodicity of a spherical coordinate becomes $\phi_1\sim \phi_1+2\pi$. And nonsingular transformation at the $z$ axis is requried ($u=\tilde{u}_-$). To prevent a conical singularity at the black hole horizon, the regularity of the black hole then requires the following periodicity $(\tau_E,x)\sim (\tau_E+\eta ,x+\zeta)$:
\ba
\eta =\dfrac{2\pi {u}_+}{u_+^2+\tilde{u}_-^2},\quad \zeta =\dfrac{2\pi \tilde{u}_-}{u_+^2+\tilde{u}_-^2}.
\ea
The above-mentioned periodicities are the same as \eqref{PER2} as we require these.

Restoring dimensions $\eta =1/(TL)$, temperature and chemical potential are given by
\ba
T=\dfrac{u_+^2+\tilde{u}_-^2}{2\pi u_+ L},\quad \mu=-\dfrac{N^x(\tilde{u}_-)}{L}=-\dfrac{i\tilde{u}_-}{u_+L}
\ea
where chemical potential is proportional to the shift \cite{Banados:1992wn} (see appendix \ref{ADM}).

We then proceed to investigate the thermodynamics of dual theory. {Total energy of the double Wick rotated metric can be computed. Considering a spatial slice of constant $t$, the radial coordinate $u$ is also fixed. The area of this slice becomes $A=\sqrt{g_{xx}} V_x=L\sqrt{u^2+u_+^2-\tilde{u}_-^2} V_x$, where $V_x=2\pi$. Using \eqref{LAP68}, the integral is computed as follows:
\ba
E=\dfrac{1}{8\pi G_3}\int dxN\sqrt{\sigma}(K-K_0)=\dfrac{L(u_+^2-\tilde{u}_-^2)}{8 G_3},
\ea
where $K_0$ is evaluated on a background with $u_+=\tilde{u}_-=0$.} By restoring dimensions ($E= \tilde{E}L$ and $P=\tilde{P}L$), one obtains
\ba
\tilde{E}=M=\dfrac{u_+^2-\tilde{u}_-^2}{8 G_3}.
\ea
This exactly agrees with energy \eqref{ENE40} of the rotating BTZ black hole.

Energy and momentum are also computed from holographic stress tensor~\cite{Fujita:2022zvz}. One needs to perform analytic continuation $\tau_E=it$. The resulting expressions become
\ba
\tilde{E}=M ,\quad \tilde{P}=\dfrac{J}{L}.
\ea

The Bekenstein-Hawking entropy is proportional to the horizon area and matches \eqref{SEN1} as follows:
\ba\label{SENn}
s=\dfrac{V_x u_+ L}{4G_3}.
\ea

Moreover, we analyze the grand canonical partition function from the classical approximation of the Euclidean path integral $Z\sim e^{-I_E[g]}$. Following the procedure outlined in section \ref{EBTZ}, we can evaluate onshell action to obtain free energy as follows:
\ba\label{EBTZ2}
&I_{E}[g]=-\dfrac{1}{16\pi G_{3}}\int d^{3}x\sqrt{g}\(R-2\L\)-\dfrac{1}{8\pi G_{3}}\int d^{2}x\sqrt{h}(K-\kappa) \nonumber \\
&=\dfrac{\eta L}{4G_{3}}(u^{2}_{\infty}-\tilde{u}_{-}^2)-\dfrac{\eta L}{4G_{3}}\left[2u^{2}_{\infty}+\(u_{+}^{2}-\tilde{u}_{-}^{2}\)\right]+\dfrac{\eta L}{4G_{3}}\left[u_{\infty}^{2}+\dfrac{1}{2}\(u_{+}^{2}-\tilde{u}_{-}^{2}\)\right] \nonumber \\
&=-\dfrac{\eta L}{8G_3}(u_+^2+\tilde{u}_-^2),
\ea
and 
\ba
F=\dfrac{I_E[g]}{\eta L}=-\dfrac{1}{8G_3}(u_+^2+\tilde{u}_-^2).
\ea
As for the second and third terms of \eqref{EBTZ2}, we have defined a modified function $A(u)=\(u^2-\tilde{u}_-^2)(u^2+u_+^2\)$, which helps us derive the results in a manner similar to the previous analysis. Note that the differences arise from $A(u)$ and the ``new" location of the horizon $\tilde{u}_{-}$. Free energy satisfies the following thermodynamic relation:
\ba
&\tilde{E}-Ts-\mu J
 \nonumber \\
&=\dfrac{u_+^2-\tilde{u}_-^2}{8G_3 }-\dfrac{u_+^2+\tilde{u}_-^2}{2\pi u_+ L}\dfrac{V_x u_+ L}{4G_3}+\dfrac{i\tilde{u}_-}{Lu_+}\dfrac{-iLu_+\tilde{u}_-}{4G_3} \nonumber \\
&=-\dfrac{u_+^2+\tilde{u}_-^2}{8 G_3 }=F,
\ea
which is consistent with free energy of the rotating BTZ \eqref{FEE43}.

{Recall that the non-zero trace of holographic stress tensor gives the Weyl anomaly. The central charge is obtained from the coefficient of the the holographic Weyl anomaly as follows~~\cite{deHaro:2000vlm, Balasubramanian:1999re}:
\ba
c=\dfrac{3L}{2G_3},
\ea
 which is identical to the one of the rotating BTZ. Recall that the central charge is independent of the mass and angular momentum. Therefore, the double Wick rotation does not affect the central charge.}

 {A crucial observation emerges from the connection between Wick rotation procedures and central charge. As demonstrated in  preceding analysis, the central charge can be universally extracted through regularization of the non-vanishing Weyl anomaly within any AdS framework. Remarkably, this derivation exhibits metric independence – the central charge determination remains invariant under specific metric choices($g_{\m\n}$), and this is  independent of the particular metric $g_{\m\n}$, as it fundamentally originates from the divergent behavior of the radial coordinate at the conformal boundary.

Recent extensions of this formalism \cite{Doi:2023zaf,Doi:2022iyj} reveal profound consequences under Wick rotation, the background spacetime turns to Euclidean case  under the wick rotation, explicitly do  on time coordinate $\t_E=i\t$,the asymptotic AdS radius acquires an imaginary character: $L_{AdS}=iL_{dS}$.This geometric transformation induces corresponding modifications in the dual CFT's central charge for dS spacetime:
\ba
c_{ds}=\dfrac{3iL}{2G_3}.
\ea
 Substituting this complex-valued central charge into entanglement entropy calculations ($S_{dS}$) for complementary subsystems $A$ and  $\bar{A}$  yields significant physical implications. The imaginary coefficient fundamentally alters the CFT's partition function $Z$, and  the pure state $|\varPsi \rangle $ of CFT is not unitary, i.e. $|\varPsi \rangle ^{\dagger}\ne \langle \varPsi |$, and the transition matrix  $\s=|\varPhi \rangle \langle \varPsi |$ replace the density matrix $\r$, similarly  the  pseudo entropy $S_{A}^{P}=-\mathrm{Tr}\left[ \sigma _A\log \sigma _A \right]$ replace entanglement entropy $S_A$, further discussion can be seen in \cite{Guo:2022sfl,Mollabashi:2020yie,Mollabashi:2021xsd,Ishiyama:2022odv}, the \cite{Doi:2023zaf,Doi:2022iyj} point out the relation between  pseudo entropy in $dS$ spacetime and timelike entropy in $AdS$ via analytic continuation of two coordinate (double wick rotation). The complex nature of $c_{dS}$  fundamentally modifies CFT properties:  complex number central charge will induce non-unitary property, in some cases represent non-Hermitian quantum system and first-order transitions \cite{Gorbenko:2018ncu,Tang:2024blm,Benedetti:2021qyk}.
}

\subsection{The propagator in the double Wick rotated BTZ}
In this section, we analyze the propagator by using the geodesic approximation in the double Wick rotated BTZ. {The double Wick rotated BTZ is a Riemannian metric in the Euclidean signature and is locally isometric to $H^3$.  
We introduce the hyperboloid ($\tilde{T
}_1,\tilde{T}_E,\tilde{X}_1,\tilde{X}_2$) where $-\tilde{T}_1^2+\tilde{T}_E^2+\tilde{X}_1+\tilde{X}_2^2=-1$.  
We parametrize the double Wick rotated BTZ as follows:
\ba\label{GEN40}
&\tilde{T}_1=\sqrt{\alpha}\cosh (\tilde{u}_{-}\tau_E+ u_+x), \nonumber \\
&\tilde{T}_E=-\sqrt{\alpha -1}\sin (\tilde{u}_- x -u_+ \tau_E), \nonumber \\
&\tilde{X}_1=-\sqrt{\alpha}\sinh (\tilde{u}_{-}\tau_E+ u_+x), \nonumber \\
&\tilde{X}_2=\sqrt{\alpha -1}\cos (\tilde{u}_- x -u_+ \tau_E).
\ea 
where $\alpha =(u^2+u_{+}^2)/(\tilde{u}_-^2+u_+^{2})$ and $u>\tilde{u}_-$. An analytic continuation was performed $\tilde{T}_E=-i\tilde{T}_2$.
$x$ is compactified with the periodicity $2\pi$. For $u_+=0$, the metric becomes a global AdS with the periodicity $x^{\prime}\sim x^{\prime}+2\pi$ ($x^{\prime}=\tilde{u}_-x$). Restricting to the hyperboloid, the double Wick rotated metric \eqref{MET2} becomes}
\ba
ds_{DW}^2=-d\tilde{T}_1^2+d\tilde{T}_E^2+d\tilde{X}_1^2+d\tilde{X}_2^2.
\ea

We analyze the chordal distance of geodesic connecting two points $(\tau_{E1}, x_1,u)$ and $(\tau_{E2},x_2,u)$ in the bulk.  Substituting \eqref{GEN40}, it becomes 
\ba
&-(\tilde{T}_1-\tilde{T}_1')^2+(\tilde{T}_E-\tilde{T}_E')^2+(\tilde{X}_1-\tilde{X}_1')^2+(\tilde{X}_2-\tilde{X}_2')^2 \nonumber \\
&=\dfrac{-2}{u_+^2+\tilde{u}_-^2}\left[u_+^2+\tilde{u}_{-}^{2}+(u^2-\tilde{u}_-^2)\cos (u_+ \Delta \tau-\tilde{u}_-\Delta x)  \right.\nonumber \\
&\left.-(u^2+u_+^2) \cosh (\tilde{u}_- \Delta \tau +u_+\Delta x) \right],
\ea
where $\Delta \tau =\tau_{E1} -\tau_{E2}$ and $\Delta x = x_1-x_2$.

By taking $u\to \infty$ limit, the chordal distance becomes
\ba
-\dfrac{2u^2}{\tilde{u}_-^2+u_+^2}(\cos (u_+\Delta \tau -\tilde{u}_-\Delta)-\cosh (\tilde{u}_-\Delta \tau +u_+\Delta x)).
\ea
This exactly agrees with the chordal distance of the rotating BTZ \eqref{CHO1}. As before, this geodesic remains spacelike for sufficiently large $u$ due to the property $\cosh (x)\ge 1$. {Due to the quotient construction, $x\sim x+2\pi$ and then $x_1$ is the equivalent class of $x_2$. Using \eqref{PRO1} and \eqref{HLO1}, the holographic correlation function becomes
\ba\label{HLO2}
&\langle \phi_{DW} (x_1)\phi_{DW} (x_2) \rangle  \nonumber \\
&=\sum_n \dfrac{(\sqrt{2}u)^{-2\Delta}({u_+^2+\tilde{u}_-^2})^{\Delta}}{(\cos (u_+ \Delta \tau_E-\tilde{u}_-(\Delta x+2\pi n))-\cosh (\tilde{u}_-\Delta \tau_E+u_+ (\Delta x +2\pi n) )))^{\Delta}}.
\ea
The formula \eqref{HLO2} agrees with \eqref{HLO1}.}

\section{Discussion}\label{DSC}
In this paper, we systematically investigate the thermodynamic property of double Wick rotated BTZ black hole and their holographic implications. 
In section \ref{DWR1}, our discussion parallels that of the Euclidean rotation BTZ but further considers the double wick rotation, it switches the time $\t_E$ and $x$ coordinate and the resultant periodicity of two coordinate mirrors that of the rotating BTZ case. The normalized thermodynamic variables yield energy and momentum densities that are identical to those obtained in the previous rotating BTZ solutions. In particular, calculation of free energy requires a redefinition of geometric parameters through the horizon radius  $\tilde{u}_-$ and a modified area functional $A(u)$. The derived  thermodynamic relations coincide with those in the rotating BTZ case. For the two point function, a similar hyperboloid transformation is implemented to analyze the chordal distance of geodesic connecting two points. This finally results in a two-point correlation function that precisely matches the expectations from rotating BTZ backgrounds. This geometric correspondence suggests deeper underlying connections between modifications of the causal structure and the propagation of holographic information propagation. 

It will be interesting to analyze the mass spectrum of fields on the Double Wick rotated metric. The metric is not a black hole but rather a global spacetime with  angular momentum in the Lorentzian frame as pointed out in \cite{Fujita:2022zvz}. It remains unclear whether or not this spacetime exhibits quasi-normal modes in the presence of imaginary angular momentum.

\section*{Acknowledgments}
We would like to thank B. S. Kim, M. A. Martin Contreras, and T. Takayanagi for their helpful discussion. 

\appendix

\section{The ADM decomposition}\label{ADM}
In this appendix, we provide a briefly review the ADM decomposition. We focus on spacelike hypersurfaces such as a constant time. These hypersurfaces are denoted by $\sigma_t$, which is spanned by $\xi^a$.  Assuming curves $\Gamma$ intersecting $\sigma_t$, we define tangent $t^{\alpha}$ as follows:
\ba\label{TAN89}
t^{\alpha}\partial_{\alpha} t=1.
\ea
A coordinate transformation is $x^{\alpha}=x^{\alpha}(t,\xi^a)$ and the following relation is obtained: 
\ba\label{TAN90}
t^{\alpha}=\Big(\dfrac{\partial x^{\alpha}}{\partial t}\Big)_{\xi^a}.
\ea
The lapse $N$ is defined in terms of the following unit normal to the hypersurface
\ba
n_{\alpha}=-N\partial_{\alpha }t.
\ea
Note that $n_{\alpha}n^{\alpha}=-1$. Because the curve $\Gamma$ is not perpendicular to $\sigma_t$, $t^{\alpha}$ is not proportional to $n^{\alpha}$. The shift $N^a$ is defined as
\ba\label{SHI92}
N^{\alpha}=t^{\alpha}-Nn^{\alpha},
\ea
where $N^{\alpha}=(\partial x^{\alpha}/\partial \xi^a)_tN^a$ and \eqref{TAN89} is satisfied.

Because of \eqref{TAN90}, \eqref{SHI92}, and the partial derivative defined as 
\ba 
dx^{\alpha}=t^{\alpha}dt+(\partial x^{\alpha}/\partial \xi^a)_t d\xi^a,
\ea 
 the metric can be expressed in the ADM form
\ba
ds^2=g_{\alpha\beta}dx^{\alpha}dx^{\beta}=-N^2 dt^2+h_{ab}(d\xi^a+N^a dt)(d\xi^b+N^b dt),
\ea
where $h_{ab}=(\partial x^{\alpha}/\partial \xi^a)_t(\partial x^{\beta}/\partial \xi^b)_t g_{\alpha\beta}$ is the induced metric on a hypersurface $\Sigma_t$ at  constant $t$. Note that $\sqrt{-g}=N\sqrt{h}$.

\subsection{Rotating BTZ black hole}
For the rotating BTZ black hole, the metric is expressed in the following ADM form:
\ba
ds^2=-N^2dt^2+u^2 (N^{x}dt+dx)^2+\dfrac{du^2}{N^2},
\ea
where the lapse and the shift vector become
\ba
N=\sqrt{-M+\dfrac{u^2}{l^2}+\dfrac{J^2}{4u^2}},\quad N^{a}=\Big(-\dfrac{J}{2u^2},0\Big).
\ea
The convention $8G_3=1$ is adopted.

\subsection{Double Wick rotated BTZ}
Performing a double Wick rotation by replacing $\tilde{t},\ \tilde{x}$ with $i x,\ i t$, respectively, the metric of the double Wick rotated BTZ is given by
\ba
ds^2=l^2 \Big(-u^2 \Big(\dfrac{u_-u_+}{u^2}dx+dt\Big)^2+\dfrac{(u^2-u_-^2)(u^2-u_+^2)}{u^2}dx^2+\dfrac{u^2 du^2}{(u^2-u_-^2)(u^2-u_+^2)} \Big).
\ea
To obtain the ADM form for the double Wick rotated BTZ, a 
 further completion of the square is required. The ADM form then becomes
\ba
ds^2=-N^2 dt^2+l^2(u^2-u_+^2-u_-^2)(dx+N^x dt )^2+\dfrac{l^2u^2 du^2}{(u^2-u_+^2)(u^2-u_-^2)},
\ea
where the lapse function and the shift vector become
\ba
N=l\sqrt{\dfrac{(u^2-u_-^2)(u^2-u_+^2)}{u^2-u_-^2-u_+^2}},\quad N^{a}=\Big(-\dfrac{u_-u_+}{u^2-u_-^2-u_+^2},0\Big).
\ea
In the Euclidean version, it is necessary to interchange $u_+$ with $\tilde{u}_-=iu_-$.

\end{document}